\documentclass[a4paper,11pt]{article}
\usepackage{pos}
\usepackage[utf8]{inputenc}
\usepackage{float}

\def\mydm{\Delta m^2_{41}}
\def\mysin{\sin^2 2\theta_{ee}}
\def\oscillationspars{$\mydm$, $\mysin$}
\def\antiparticle{\tilde}

\def\clsmethod{CL$_s$}

\title{New results from the DANSS experiment.}

\author*[a]{Mikhail Danilov}

\affiliation[a]{Lebedev Physical Institute of the Russian Academy of Sciences,\\ 53 Leninskiy Prospekt, Moscow, 119991, Russia.\\On behalf of the DANSS Collaboration}

\emailAdd{danilov@lebedev.ru}

\abstract{

We present new preliminary DANSS results based on 3.5 million Inverse Beta Decay (IBD) events collected at 3 distances (10.9~m, 11.9~m, and 12.9~m) from the detector center to the reactor core center. The detector position is changed typically 3 times a week. Therefore many systematic uncertainties are canceled out.  A new analysis that uses information about relative IBD counting rates in addition to changes in positron energy spectra shapes is employed. The excluded area covers a very interesting range of the sterile neutrino parameters up to $\mysin < 0.008$ in the most sensitive region.  No statistically significant evidence for sterile neutrinos is observed. 
The  significance of the best-fit point in the 4$\nu$ case is 1.5$\sigma$.
}

\FullConference{%
  40th International Conference on High Energy physics - ICHEP2020\\
  July 28 - August 6, 2020\\
  Prague, Czech Republic (virtual meeting)
}

\begin{document}
\maketitle

\section{Introduction}

Sterile neutrinos appear naturally in many extensions of the Standard Model (SM). 
Moreover there are several experimental indications of their existence.
The deficit of $\nu_e$ in the calibration of the SAGE and GALEX experimets with radioactive sources~\cite{SAGE, GALEX} (``Galium Anomaly''(GA)) and the deficit in reactor $\antiparticle\nu_e$ fluxes~\cite{Mueller} (``Reactor Antineutrino Anomaly''(RAA)) can be explained by active-sterile neutrino oscillations \cite{Ga,Mention2011}. 
The MiniBooNE collaboration presented recently  a 4.8$\sigma$ evidence for ($\antiparticle\nu_e$)$\nu_e$ appearance in the muon (anti)neutrino beams~\cite{MiniBooNE2} confirming earlier LSND results. 
The Neutrino-4 experiment claimed an observation of $\antiparticle\nu_e$ oscillations to sterile neutrinos although the significance of the result is only 3.2$\sigma$\cite{Neutrino4-2019, Neutrino4-2020} and there are concerns about the validity of their analysis~\cite{DanSkr}.

The survival probability of reactor $\antiparticle\nu_e$ at very short distances in the 4$\nu$ mixing scenario (3 active and 1 sterile neutrino) is given by the formula:

\begin{equation}
\label{eqosc}
1-\mysin \sin^2\left(\frac {1.27\mydm [\mathrm{eV}^2] L[\mathrm m]}{E_\nu [\mathrm{MeV}]}\right),
\end{equation}
where $\mysin$ is the mixing parameter, $\mydm = m_4^2 - m_1^2$ is the difference in the squared masses of neutrino
mass states, $L$ is the distance between production and detection points and $E_{\nu}$ is the $\antiparticle\nu_e$ energy.
 The Inverse Beta Decay (IBD) reaction 
$\antiparticle{\nu}_e + p \rightarrow e^+ + n$ 
is used to detect $\antiparticle\nu_e$.
\section{Energy calibration and backgrounds}
The DANSS detector~\cite{DANSS} is located on a movable platform under 
an industrial 3.1~GW$_{th}$ reactor of the Kalininskaya NPP in Russia.
 The DANSS detector consists of 2500 one meter long scintillator strips with Gd-loaded surface coating. Strips in neighbor layers are orthogonal. This allows a quasi-3D reconstruction of events.  Each strip is readout with 
3 wavelength-shifting fibers placed in grooves along the strip. The central
fiber is read out by a SiPM and the two side fibers from 50
strips are bundled together and readout by a  PMT. The scintillator detector is
surrounded by a multi-layer passive shielding of copper, lead and borated 
polyethylene. Double layers of scintillator counters provide active shielding of the detector from all sides with exception of the bottom one.
The high granularity of the detector allows the reconstruction of the positron track. Its kinetic energy is used in the analysis without adding energies of annihilation gammas that suffer from a nonlinear energy response in all experiments. 
Positron energy is 1.02~MeV smaller than the prompt energy used in other experiments. 
The $\antiparticle\nu_e$ energy is 
$E_{\antiparticle\nu} \approx E_{e^+} + 1.8~\mathrm{MeV}$. 

The detector energy scale is anchored by the electron energy spectrum from $^{12}$B decays. Electron signals are very similar to positron signals apart from the annihilation gammas that are not used for the energy determination. In both cases $e^+$ or $e^-$ produce high ionization density only at the end of the track. Thus saturation effects that are difficult to describe are relatively small in both cases. Therefore the non-linearity of the detector response is expected to be small and reasonably well described by the Monte Carlo (MC) simulations.  

 Michel electrons from cosmic muons decays provide a cross-check of the energy calibration. The total
energy in the delayed events following stopped muons is reconstructed. 
The best agreement with MC is achieved with the MC energy scale shifted by -1.7\%.
This is natural since the $e^+$ annihilation following  $\mu^+$ decays leads to many soft electrons for which saturation effects are important. The energy scale determined using  $^{22}$Na, $^{60}$Co, and $^{248}$Cm radioactive sources is also shifted from the $^{12}$B scale by -1.5\%, -1.0\%, and -0.5\% correspondingly. The shift is larger for sources with larger number of soft electrons. Thus our MC describes reasonably even processes with many soft electrons for which saturation effects are important. One can expect even better agreement in case of  $^{12}$B decays with only one soft electron. Nevertheless we keep a conservative estimate of 2\% for the energy scale uncertainty in the analysis.

 Observed energy resolution for different calibration sources is slightly worse than MC predictions (33\% instead of 31\% at 1 MeV). Therefore additional smearing is added to the MC predictions ($\sigma_{additional}/E = 12\%/\sqrt{E} \oplus 4\%$). With this correction MC describes well all calibration sources.

The accidental coincidence background is the largest background in the experiment. It is calculated without any model dependence using  time intervals for the neutron signal shifted back in time with respect to the $e^+$ signal. Sixteen intervals are used in order to reduce the statistical error to a negligible level in comparison with the IBD signal. Nevertheless subtraction of the accidental background increases statistical errors of the IBD $e^+$ spectrum. Therefore several cuts are used to suppress this background~\cite{DANSSdata}.
In the present analysis the lower cut on the neutron signal energy and the upper cut on the distance between $e^+$ and neutron positions depend on the $e^+$ energy. For smaller $e^+$ energies where background is larger these cuts are more restrictive. The energy cut changes from 4~MeV to 1.5~MeV and the distance cut changes from 33~cm to 43~cm.
With such optimized cuts the accidental background is only 13.8\% of the IBD signal at the top detector position (this position is used for other background estimates below).

The DANSS detector is placed under the reactor that provides an overburden of about 50 mwe. 
Such shielding removes all cosmic neutrons apart of those produced by muons in the detector hall walls. This background is estimated by a linear extrapolation from the (10-16)~MeV range of the $e^+$ spectrum to  a (1.5-6)~MeV range used in the fits. It amounts to 0.4\% of the IBD signal. 
Cosmic muons can produce one or several neutrons in the detector shielding. These neutrons can mimic the IBD process producing a pair of signals correlated in time. The shape of this background is measured using events tagged by the muon veto system which removes the major part of the muon induced background. The inefficiency of this system is determined using the reactor-off data. As a result the muon induced background constitutes only 1.5\% of the IBD signal.
The background from neighbor reactors has a well known shape and constitutes 0.6\% of the IBD signal. All other backgrounds are negligible. If we do not consider backgrounds with well known shape and normalization the    
signal/background ratio exceeds 50. 

\section{Positron spectra}
\begin{figure}[h]
\begin{center}
\begin{tabular}{cc}
\includegraphics[width=0.54\textwidth]{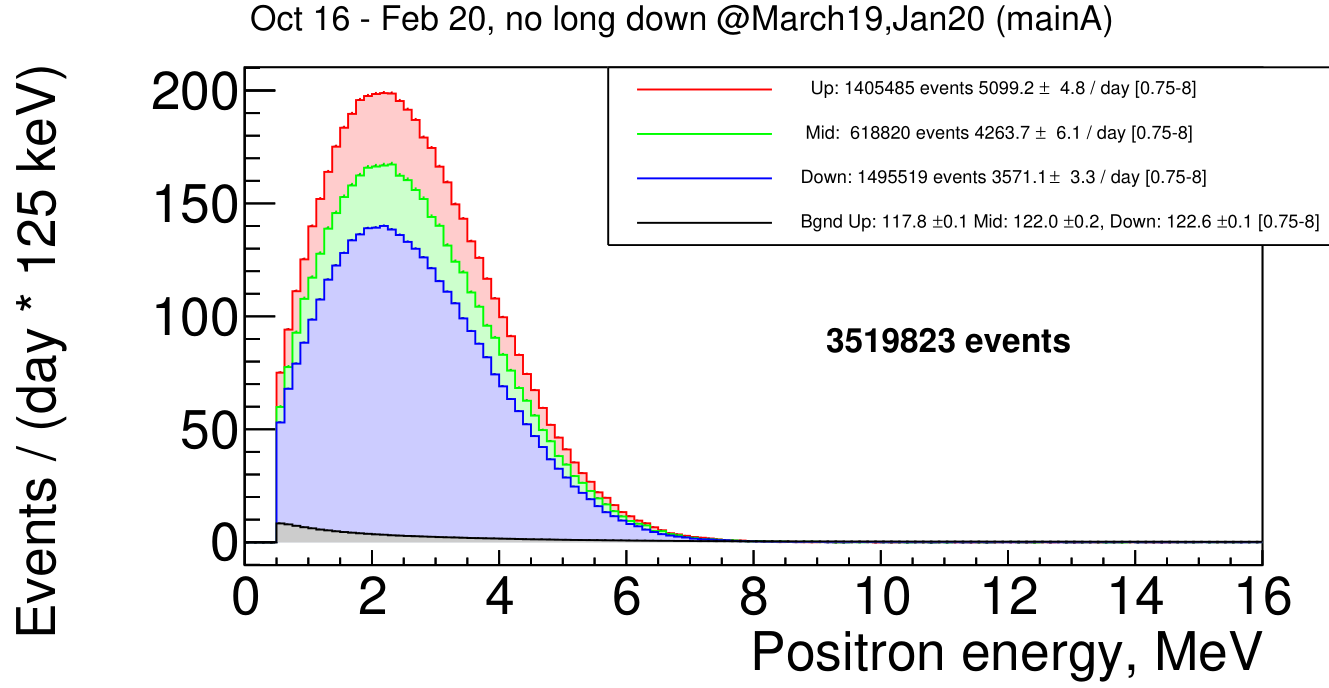} &
\includegraphics[width=0.40\textwidth]{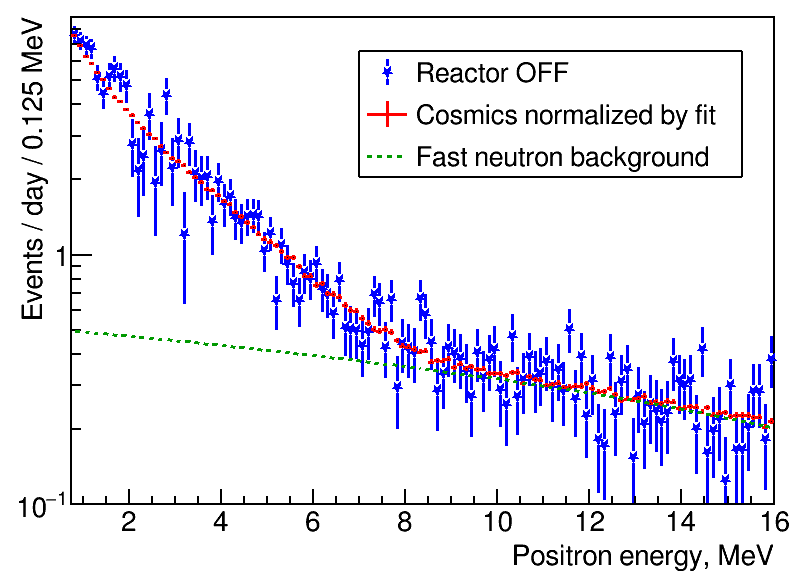}
\end{tabular}
\end{center}
\vspace{-0.6cm}
\caption{\label{fig:spect} Positron spectra at 3 detector positions (left) and $e^+$ spectrum during reactor-off periods (right).}
\vspace{-0.2cm}
\end{figure}

The positron spectra are shown in figure~\ref{fig:spect}.
 The IBD rate at the top detector position exceeds 5 thousand events per
day with the background of about 90 events per day after subtraction of  30 events per day from neighbor reactors.

\begin{figure}[h]
\begin{center}
\begin{tabular}{cc}
\includegraphics[width=0.47\textwidth]{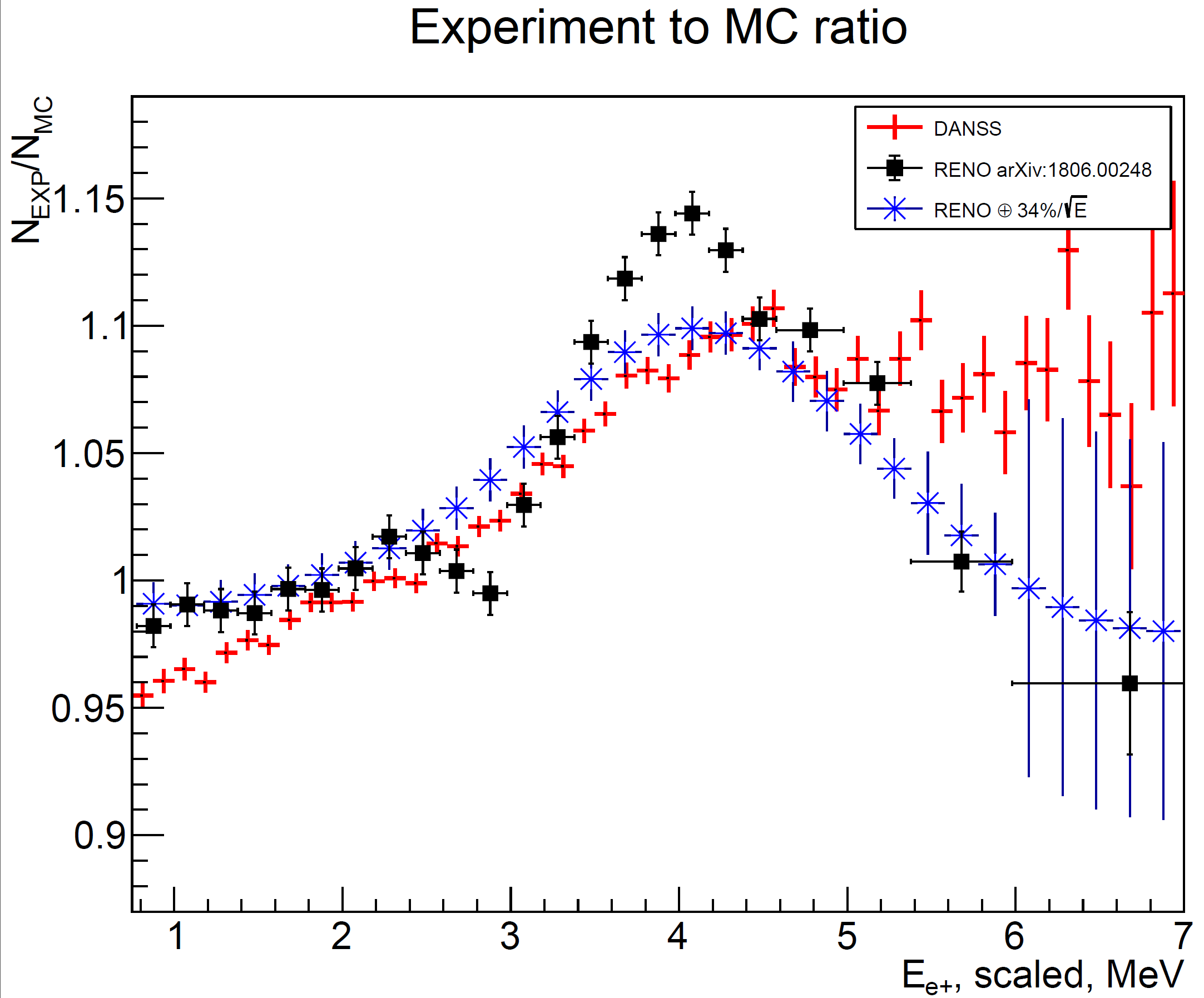} &
\includegraphics[width=0.47\textwidth]{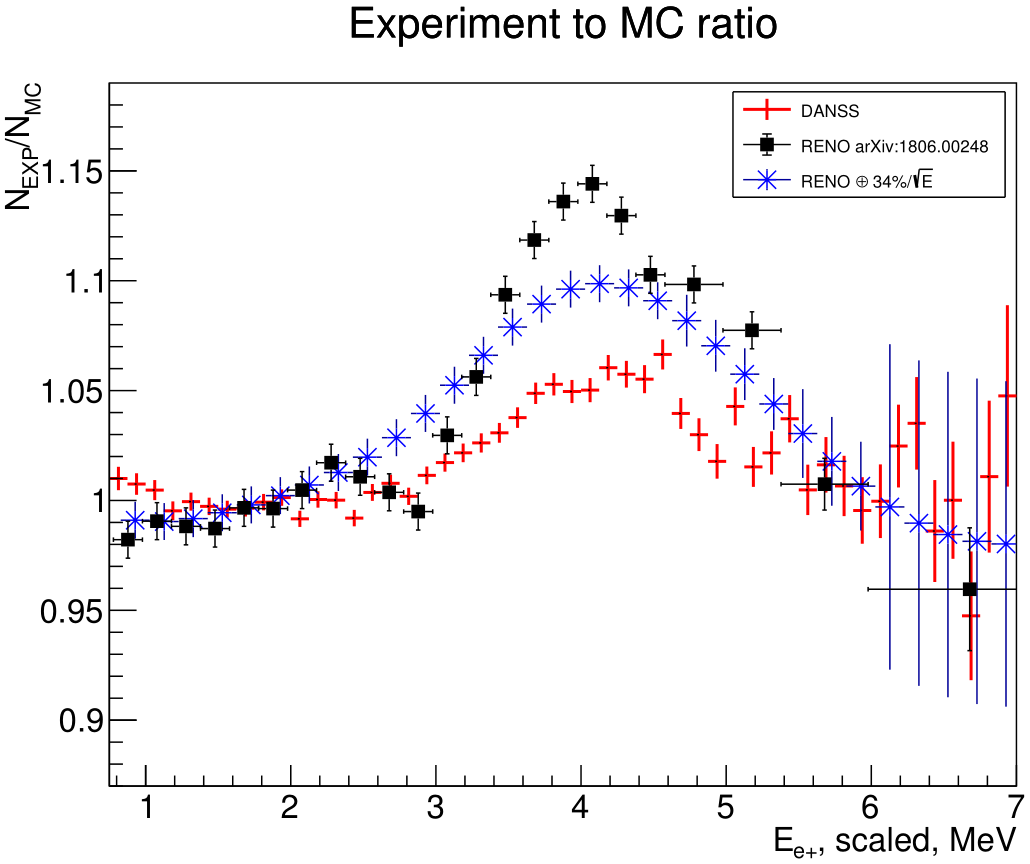}
\end{tabular}
\end{center}
\vspace{-0.6cm}
\caption{\label{fig:bump} Experiment to MC ratio for $e^+$ spectrum without (left) and with a -50~keV shift in energy (right).}
\vspace{-0.2cm}
\end{figure}

A ratio of the measured $e^+$ spectrum to the MC simulated Huber-Mueller spectrum~\cite{Huber,Mueller} is shown in figure~\ref{fig:bump} (left).
RENO results~\cite{RENO} are shown
for comparison shifted by 1.02~MeV to correct for two 511~keV annihilation gammas. The RENO spectrum smeared by the DANSS
energy resolution is also shown. The best agreement between our measurements and MC
in the range $1.5 - 3.0$~MeV is obtained if data are shifted by -50~keV (right figure). 
The nature of this shift (if it exists) is unknown to us. Therefore we can not claim observation of the bump in our data although some structure in this region is obviously seen.  
The search for sterile neutrinos does not depend on the exact shape of the $e^+$ spectrum since we consider only ratios of $e^+$ spectra at the different positions from the reactor core.  

\section{Search for sterile neutrinos.}

\begin{figure}[h]
\includegraphics[width=0.45\textwidth]{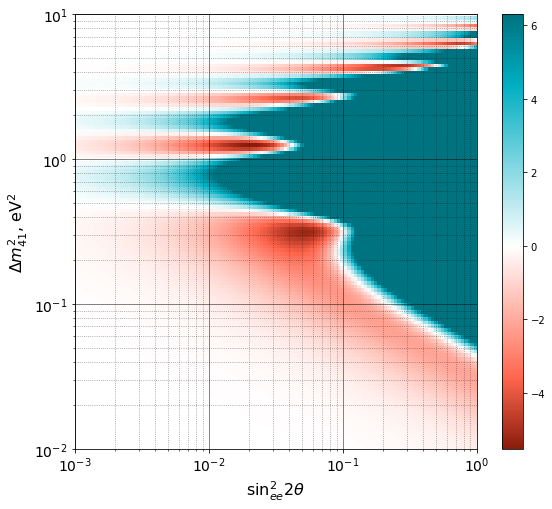}\hspace{2pc}%
\begin{minipage}[b]{0.45\textwidth}\caption{\label{fig:delta_chi_map} Distribution of $\Delta \chi^2 = \chi^2_{4\nu} - \chi^2_{3\nu}$ for the parameter space \oscillationspars\ (color online). Blue color indicates areas with $\chi^2_{4\nu} > \chi^2_{3\nu}$, red color indicates areas with $\chi^2_{4\nu} < \chi^2_{3\nu}$. The color axis is limited, $\Delta \chi^2_{max} = \chi^2_{best 4\nu} + 11.83$. The value 11.83 corresponds to the exclusion at $3\sigma$ level in case of a $\chi^2$ distribution with 2 d.o.f. In case of sterile neutrino searches $\Delta \chi^2$ statistics doesn't follow the $\chi^2$ distribution with 2 d.o.f., so here we use it just for illustration. Exclusion areas are calculated with a Gaussian \clsmethod\ method.}
\end{minipage}
\end{figure}

\begin{figure}[h]
\centering
\begin{minipage}[t]{0.53\linewidth}
\includegraphics[width=\linewidth]{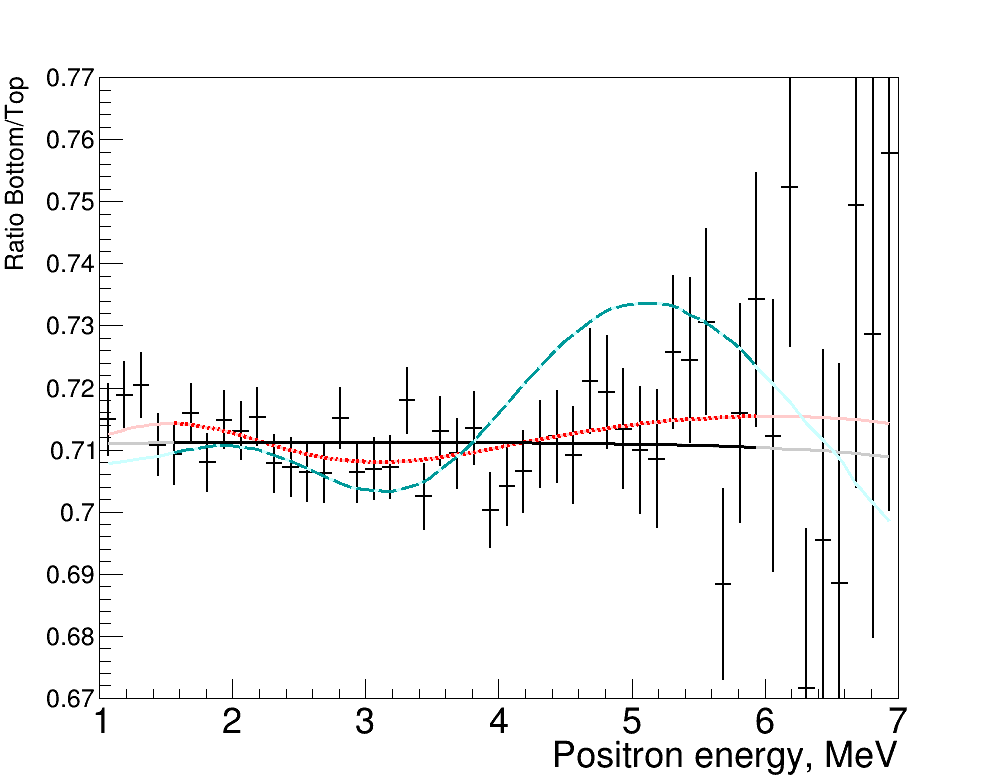}
 \caption{\label{fig:du} Ratio of $e^+$ energy spectra measured at the bottom and top detector positions (statistical errors only). The solid curve is the prediction for 3$\nu$ case, the dotted curve corresponds to the best fit in the $4\nu$ mixing scenario ($\mysin = 0.02$, $\mydm = 1.3~\rm{eV}^2$), the dashed curve is the expectation for the optimum point from the RAA and GA fit \cite{Mention2011} ($\mysin=0.14$, $\mydm = 2.3~\rm{eV}^2$,$\Delta \chi^2 =68$).}
  \end{minipage} 
  \hfil
\begin{minipage}[t]{0.41\linewidth}
\includegraphics[width=\linewidth]{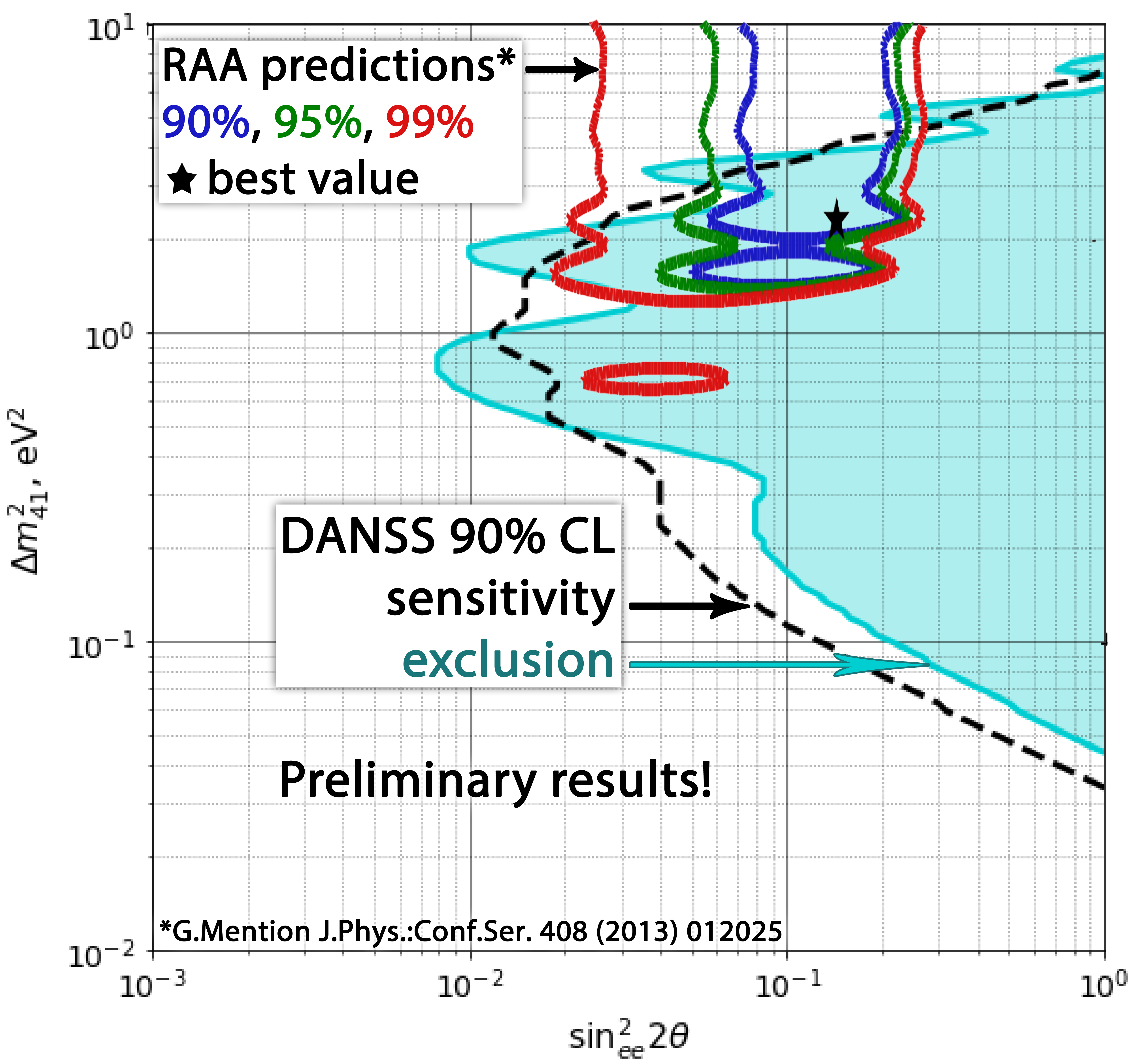}
 \caption{\label{fig:cls} Exclusion area at 90\% C.L. obtained with the Gaussian \clsmethod\ method (filled area) and 90\% C.L. sensitivity contour (dashed line). Expected regions from RAA and GA are also shown.}
 \end{minipage}\hspace{2pc}%
 \end{figure}

For a grid of points in the \oscillationspars\ plane predictions for the $e^+$ spectra at the different detector positions are calculated using the full MK detector response and the Huber-Mueller model~\cite{Huber,Mueller}. Results do not depend on the choice of the model since we consider only the ratios of $e^+$ spectra at different detector positions. The predictions are compared with the data using  $\Delta \chi^2 = \chi^2_{4\nu} - \chi^2_{3\nu}$ that includes systematic uncertainties treated as nuisance parameters. 
Systematic uncertainties include 2\% in energy scale, 50~keV shift in energy, 25\% in cosmic background, 30\% in fast neutron background, 25\% in additional smearing of the energy resolution, 5~cm in distance to the fuel burning center, and 0.2\% in relative detector efficiencies at different distances.
Relative IBD counting rates at different detector positions are used now for $\chi^2$ calculations~\cite{RelEfficiency} contrary to the previous analysis in which only $e^+$ spectra shapes were compared~\cite{DANSSdata}.  The corresponding uncertainties are included as penalty terms for nuisance parameters into the test statistics. The middle detector position is used now in the $\chi^2$ calculations. Previously we used it only for cross-checks. The obtained $\Delta \chi^2 = \chi^2_{4\nu} - \chi^2_{3\nu}$ distribution is shown in figure~\ref{fig:delta_chi_map}.
The best fit point in the parameter space is $\mydm = 1.3$eV$^ 2$, $\mysin=0.02$ and $\Delta \chi^2 = -5.5$. The statistical significance of such a difference is determined by the Feldman-Cousins method~\cite{FeldmanCousins} to be 1.5~$\sigma$. Figure~\ref{fig:du} shows the $Bottom/Top$ spectra ratio.
The exclusion area calculated using a Gaussian \clsmethod\ method~\cite{CLS} is shown in figure~\ref{fig:cls}. It covers a very large and probably the most interesting fraction of the expectations based on the 
RAA and 
GA~\cite{Mention2011}. The best fit point of the RAA+GA is excluded at more than 5$\sigma$ level using the Gaussian \clsmethod\ method that usually gives more conservative results than the Feldman-Cousins method.  

\section{Conclusions}

Larger statistics and inclusion of the relative IBD counting rates at different detector positions into analysis allowed us to extend the exclusion area in the \oscillationspars\ parameter space and exclude a large and the most interesting fraction of the RAA and GA predictions. The best fit point of RAA and GA is excluded with more than 5$\sigma$ significance.
No statistically significant evidence for sterile neutrinos is observed. The significance of the best-fit point is 1.5$\sigma$.

\section*{Acknowledgments}
 
The collaboration appreciates the permanent assistance of the KNPP administration
and Radiation Safety Department staff.
This work is supported by the Ministry of science and higher education of
Russian Federation under the contract 13.1902.21.0005.

\end{document}